\documentclass[prl,aps,amsmath,amssymb,reprint,floatfix,superscriptaddress]{revtex4-2}

\pdfoutput=1
\usepackage{xcolor}
\usepackage{graphicx}
\usepackage{textcomp}
\usepackage[utf8]{inputenc}
\usepackage[T1]{fontenc}
\usepackage[colorlinks=true,citecolor=blue,urlcolor=blue]{hyperref}
\usepackage{etoolbox} 
\usepackage{bm} 
\usepackage[normalem]{ulem}
\usepackage{physics}
\usepackage{hyperref}

\def\eqq#1{Eq.~(\ref{#1})}

\def\eq#1{(\ref{#1})}

\def\c#1{~\cite{#1}}

\def\dd{{\rm d}}

\def\av#1{\langle #1 \rangle}

\def\beq{\begin{equation}}
\def\eeq{\end{equation}}
\def\bea{\begin{eqnarray}}
\def\eea{\end{eqnarray}}

\def\kB{k_{\rm B}}
\def\tf{t_{\rm f}}
\def\kt{\kB T}

\def\e{{\rm e}}

\begin{document}

\title{Adding noise and scaling forces to speed up the Langevin clock}

\author{Prithviraj Basak}
\email{pbasak@sfu.ca}
\affiliation{Department of Physics, Simon Fraser University, Burnaby, British Columbia V5A 1S6, Canada}

\author{Stephen Whitelam}
\email{swhitelam@lbl.gov}
\affiliation{Molecular Foundry, Lawrence Berkeley National Laboratory, 1 Cyclotron Road, Berkeley, CA 94720, USA}

\author{John Bechhoefer}
\email{johnb@sfu.ca}
\affiliation{Department of Physics, Simon Fraser University, Burnaby, British Columbia V5A 1S6, Canada}

\begin{abstract}
Using experiments on a colloidal particle trapped in an optical tweezer, we confirm a recent proposal to increase the effective mobility or clock rate of systems described by Langevin dynamics, by simultaneously scaling deterministic forces and adding external noise. A corollary, which we also confirm experimentally, is that a system driven out of equilibrium by a time-dependent protocol can remain closer to thermal equilibrium. As an application, we demonstrate more precise recovery of free-energy differences from nonequilibrium work relations.  Langevin clock rescaling provides a general strategy for accelerating calculations in the emerging field of thermodynamic computing, which uses stochastic devices governed by Langevin dynamics to do low-energy calculations. 

\end{abstract}

\maketitle

The overdamped Langevin equation describes the time evolution of a wide range of mesoscopic systems for which thermal fluctuations play an essential role~\cite{Ermak1978,lemons1997, Kampen2007, Coffey2017}. Such systems include biomolecular complexes in living cells, and single-molecule experiments involving optical tweezers and atomic force microscopes~\cite{bustamante2005,Ritort2006,Greenleaf2007,Miller2017}. 
In a Langevin system, the dynamics are governed by deterministic forces, such as those from a potential, the thermal kicks from the surrounding bath, and the mobility parameter. Since the mobility sets the basic relaxation time, increasing it would accelerate the dynamics of a broad class of Langevin systems.  In practice, however the mobility of a system is fixed by the physical properties of the system and its environment,  for example by the viscosity of the medium or the size of a colloidal particle. The range of these physical properties limits the range over which the system's mobility can vary.

Here, we take an alternate approach inspired by a recent work to \emph{effectively} increase the mobility of an arbitrary Langevin device: by scaling the deterministic force and adding external noise in the correct proportion, we obtain a new Langevin equation that looks like the original equation but with 
rescaled mobility~\cite{Whitelam2025a}. The result is an effective speed-up of the dynamics of the system, or a rescaling of the Langevin clock. 

    \begin{figure}[ht]
        \centering
        \includegraphics[width=\linewidth]{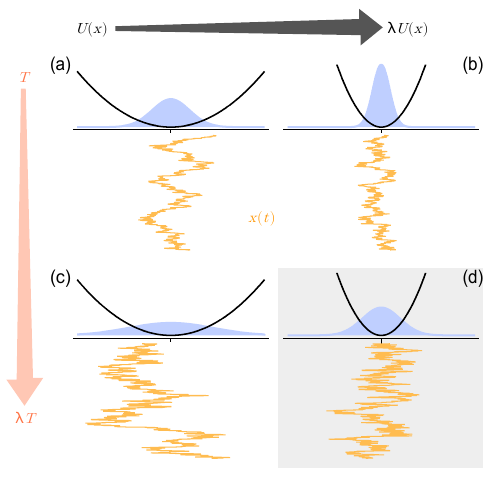}
        \caption{ Schematic of speed-up concept. (a) In a harmonic potential $U(x)$, a particle has a Gaussian position distribution $p(x)$ (blue shading) that is compiled from individual trajectories of particle position, $x(t)$ (orange trace). (b, c) From left to right, the potential $U(x)$ is scaled by a factor $\lambda$.  From top to bottom, the temperature $T$ is scaled by $\lambda$. Scaling either $U(x)$ or $T$  alters the position distribution. (d) Simultaneously scaling $U(x)$ and $T$ leaves $p(x)$ unchanged, while speeding up dynamics.}
        \label{fig_scheme}
    \end{figure}

Using this strategy, we demonstrate experimentally a speed-up of Langevin dynamics, for the paradigmatic example of a single colloidal particle in an optical trap, by more than an order of magnitude. The particle is confined in a harmonic potential $U$ and undergoes Brownian motion from the thermal kicks at room temperature $T$ (Fig.~\ref{fig_scheme}(a)). We can independently scale the potential $U \to U_\lambda=\lambda U$ by increasing the laser power and raise the effective temperature $T\to T_\lambda = \lambda T$ by adding random fluctuations to the trap position (the bath temperature $T$ remains essentially unchanged).  Here, $\lambda\ge 1$ is the Langevin-clock scaling parameter. When scaling either the effective temperature {\em or} the potential (Fig.~\ref{fig_scheme}(b), (c)), the stationary distribution of the system changes. Upon simultaneous scaling of both the effective temperature {\em and} the potential by the same  factor $\lambda$, the stationary distribution is unaltered (Fig.~\ref{fig_scheme}(d)), but the effective mobility parameter increases (note that the bare mobility parameter, which depends on the viscosity of water and the size of the particle, is unchanged)~\cite{Whitelam2025a}. One consequence of this effective increase of mobility is that the particle remains closer to equilibrium during a nonequilibrium process, such as when pulled through a fluid at constant speed. As a result, the effective dissipation is reduced, and we can more precisely measure the free-energy change of the transformation using nonequilibrium work relations~\cite{Whitelam2025, Jarzynski1997, Crooks1999}.

This change of effective temperature is achieved by the controlled addition of noise. Adding noise to single-particle experiments~\cite{Martinez2013, Chupeau2018, Saha2023, DiBello2024, Message2025} or thermodynamic computers~\cite{aifer2024} represents a powerful method of control. For instance, Chupeau et al. demonstrate that time-varying noise can accelerate single-molecule relaxation dynamics, reducing the relaxation time to thermal equilibrium under ambient conditions~\cite{Chupeau2018}. The strategy we use is related but distinct: by controlling potential energy and noise in the correct proportion, the system also evolves faster than under ambient conditions but is statistically indistinguishable from a time-rescaled version of the original experiment.

This strategy applies in principle to any Langevin system.   In particular, it could be useful in the emerging field of thermodynamic computing, in which low-energy computation is done by physical devices whose degrees of freedom evolve in time according to the Langevin equation~\cite{Conte2019,Wimsatt2021, Boyd2022,aifer2024}. Such devices have been shown to perform computations such as high-throughput sampling in sparse Ising machines, circuit implementations of matrix inversion, and generative modeling akin to diffusion models~\cite{Aadit2022,Melanson2025,Duffield2025,Whitelam2026}. These computations are executed by stochastic relaxation toward a target distribution, so that accelerating Langevin dynamics would increase throughput in thermodynamic hardware~\cite{Coles2023,Wolpert2024,Melanson2025,Whitelam2026}. The time scale of operation (mobility) of a thermodynamic computer is set by the particulars of the device, with optical, mechanical, electronic, and Josephson junction devices having mobilities that span about nine orders of magnitude ($\text{s}^{-1}$ to $\text{ns}^{-1}$)~\cite{Jun2014,Saira2020,Dago2021,Ray2023,Pratt2025}.  The present approach, by contrast, provides a general way to increase the clock speed of a thermodynamic computer for a given type of device.
  
{\em Langevin clock rescaling}---Consider a classical fluctuating degree of freedom $x$~\footnote{The extension to $N$ degrees of freedom is straightforward~\cite{Whitelam2025}} in contact with a thermal reservoir at temperature $T$. The degree of freedom experiences a potential $U(x,t)$ and evolves in time according to the Langevin equation 
\beq
\label{orig}
\dot{x} = -\mu \frac{\partial U(x,t)}{\partial x} + \sqrt{2 \kt \mu} \, \eta_0(t).
\eeq
Here $\mu$ is the mobility, and $\eta_0(t)$ is a Gaussian white-noise source with correlations $\av{\eta_0(t)}=0$ and $\av{\eta_0(t) \eta_0(t')}=\delta(t-t')$. We consider the evolution of Eq.~\eq{orig} on the time interval $t \in [0,\tf]$. This is our original, unmodified system.

We now scale the potential by a factor of $\lambda>1$ and add to the system an independent 
source of Gaussian white noise, $\sqrt{2 \kt \mu (\lambda-1)} \eta_1(t)$, with $\av{\eta_1(t)}=0$, $\av{\eta_1(t) \eta_1(t')}=\delta(t-t')$, and $\av{\eta_0(t) \eta_1(t')} = 0$. 
The Langevin equation describing this modified system is
\bea
\label{mod}
\dot{x} = -\mu \frac{\partial \lambda U(x,t)}{\partial x} &+& \sqrt{2 \kt \mu} \, \eta_0(t) \nonumber \\
&+& \sqrt{2 \kt \mu (\lambda-1)} \, \eta_1(t).
\eea
The two Gaussian white noise terms can be written as a single Gaussian white noise of variance $2 \kt \lambda \mu$, and the effective equation of motion becomes
\beq
\label{eff}
\dot{x} = -\mu_\lambda \frac{\partial U(x,t)}{\partial x} + \sqrt{2 \kt \mu_\lambda} \, \eta(t),
\eeq
with $\mu_\lambda \equiv \lambda \mu\geq\mu$ and where $\eta(t)$ also has white-noise correlations. Equation~\eqref{eff} shows that the modified system is equivalent to a version of the original system, \eqq{orig}, with a larger mobility parameter~\cite{Whitelam2025}. 

  \begin{figure*}[ht]
        \centering
        \includegraphics[]{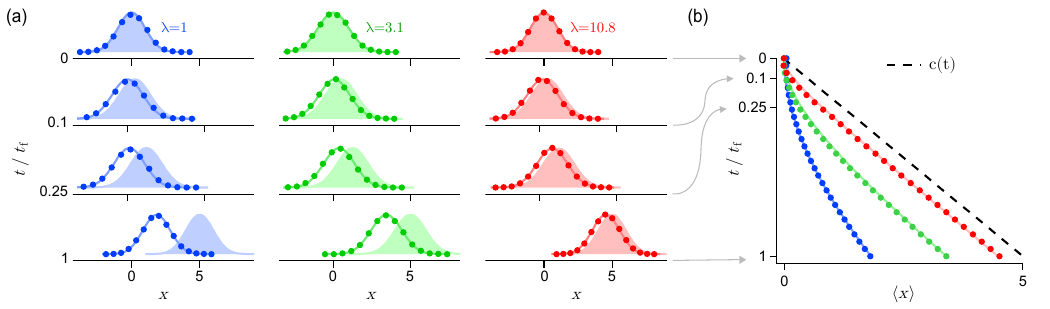}
        \caption{Experimental demonstration of the Langevin clock rescaling.
        (a) Time evolution of the measured particle position distribution $p(x,t)$ (markers) under the trap motion $c(t)$.  The shaded distributions represent the equilibrium distribution corresponding to the trap position. The solid lines are predictions for our nonequilibrium transformation whose mean is given by Eq.~\ref{eqn:x(t)}. At $\lambda=1$, the distribution stays farther from the trap center at the end of the protocol.  But at higher $\lambda$, the distribution stays closer to the equilibrium distribution.  (b) The trap center $c(t)$ moves at constant velocity over a distance $5\sigma_x$, within one relaxation time $t_\textrm{f}=\tau_\textrm{r}$.  At $\lambda=1$, the ensemble-averaged position of the particle $\langle x_n\rangle$ remains farther from the trap position and does not reach a steady state.  At $\lambda=10.8$, the mean position $\langle x_n\rangle$ relaxes faster and reaches a steady state much closer to the trap position.  The solid curves are the predictions $\langle \tilde{x}(\tilde{t})\rangle$ based on Eq.~\ref{eqn:x(t)}. }
        \label{fig_px_xt}
    \end{figure*}    
  
    \emph{Experiment---} Our experimental overdamped Langevin system involves a $1.5$\,\textmu m silica bead suspended in water and confined using a feedback-optical tweezer~\cite{Kumar2018,Bryan2022} (Sec.~I in Supplementary Material~\cite{SI}). The harmonic trap position is controlled by an acousto-optic deflector (AOD), which adjusts the laser beam angle to displace the trap center.  A linear deflection-voltage in the AOD translates the trap at a constant velocity.  The same device also modulates laser power, allowing independent control of trap stiffness. We acquire data and run the control loop using an FPGA at a sampling time $\Delta t = 20$ \textmu s. Thus, the experimentally measured position $x_n$ is sampled at discrete time $t_n$, where $t_n - t_{n-1} = \Delta t$.
    
    In our optical tweezer, a Brownian particle suspended in water experiences a harmonic potential,
    \begin{equation}
        U_\lambda(x,c(t)) = \frac{\kappa_\lambda}{2}\,
        \left( x-c(t) \right)^{2},
    \end{equation}
    where $\kappa_\lambda$ is the trap stiffness and $c(t)$ the trap center. To drive the system out of equilibrium, the trap is translated at constant velocity according to
    \begin{equation}
        c(t) = c_\textrm{f}\,\frac{t}{t_\textrm{f}}.
    \end{equation}
    Here the total displacement is $c_\textrm{f}=5\sigma_x$, where $\sigma_x^2 = \kB T/ \kappa_1$, is the position variance, and the protocol duration $t_\textrm{f}$  equals the relaxation time $\tau_\mathrm{r}= 1/(\kappa_1\mu)$. In our discrete time control system, the trap center translation also follows discrete steps $c_n = c_\textrm{f}\tfrac{t_n}{t_\textrm{f}}$ (Sec.~VI in Supplementary Material~\cite{SI}).

    We first determine a baseline stiffness $\kappa_1$ at which the bead remains stably trapped, and we treat it as our original unscaled system. After that, the scaling approach is realized in two steps: first the potential is altered, and then the effective temperature. Increasing the laser power raises the stiffness to $\kappa_\lambda$, from which the experimentally implemented scaling factor is obtained as
    \begin{equation}
            \lambda = \frac{\kappa_\lambda}{\kappa_1}.
    \end{equation}
    
    For each $\kappa_\lambda$, we implement the effective temperature $T_\lambda$ by adding noise. The extra noise is introduced at discrete time $t_n$ by applying Gaussian random displacements $\sigma_c \eta_n$ to the trap center, similar to~\cite{Chupeau2018}.
    Here, $\eta_n$ is a pseudorandom number drawn from a Gaussian random-number generator with 
    $\av{\eta_n}=0$, $\av{\eta_n \eta_m}=\delta_{nm}$.  
    The noise force is generated as $\kappa_\lambda\sigma_c \eta_n$. We choose the variance of the trap center displacement as, 
    \begin{equation}
            \sigma_c^2 = (\lambda - 1)\frac{2\mu}{\kappa_\lambda^2 \Delta t},
    \end{equation}
    so that the effective temperature increases to $T_\lambda = \lambda T$.

    \emph{Results---}We realized the constant-velocity protocol at different Langevin-clock rescaling parameter values $\lambda$, ranging from 1 to 10.8. Figure~\ref{fig_px_xt}(a) shows the time evolution of the position distribution $p(x,t)$ for three values of $\lambda$. As $\lambda$ increases, the position distribution relaxes towards equilibrium more rapidly, consistent with theoretical predictions of sped up overdamped Langevin relaxation. The ensemble-averaged position $\langle x_n\rangle = \tfrac{1}{N} \sum_{i=1}^{N} x_{n,i}$ likewise approaches equilibrium faster at larger $\lambda$ (Fig.~\ref{fig_px_xt}(b)). Here, $N$ is the number of trials. The predicted mean position~\cite{Mazonka1999} plotted as the solid curves in Fig.~\ref{fig_px_xt} follows
    \begin{equation}
            \langle \tilde{x}(\tilde{t})\rangle = \tilde{c}(\tilde{t}) + \frac{\tilde{c}_\textrm{f}}{\lambda \tilde{t}_\textrm{f}} \left(e^{{-\lambda \tilde{t}}} - 1 \right).
        \label{eqn:x(t)}
    \end{equation}
      The dimensionless trap displacement and protocol time are $\tilde{c}_\textrm{f} = c_\textrm{f}/\sigma_x, \tilde{t}_\textrm{f} = t_\textrm{f}/\tau_\textrm{r}$, and $\tilde{x}(t) = x\sigma_x, \tilde{t}=t/\tau_\textrm{r}$.  We also see that at the highest scaling factor the dynamics reaches a nonequilibrium steady state while the unscaled dynamics does not (Fig.~\ref{fig_px_xt}(b)).  The asymptotic separation between the bead's position and the trap position, $\langle \tilde{x}(\tilde{t})\rangle - \tilde{c}(\tilde{t}) \propto 1/\lambda$, reduces with increasing scaling factor. 
     
     As evident from Eq.~\ref{eqn:x(t)}, we can also interpret the effect of scaled dynamics as if the protocol were applied for $\lambda$ times longer. These observations establish that our $\lambda$-scaling indeed speeds up the basic relaxation rate of our system. Consequently, during a transformation at fixed rate, the $\lambda$-scaled system remains closer to equilibrium than does the original system. We predict\c{Whitelam2025} that this effect should lead to reduced effective dissipation (scaled by the reciprocal temperature $\beta_\lambda = \tfrac{1}{k_\textrm{B}T_\lambda}$) and therefore work statistics that lend themselves to more accurate application of nonequilibrium work relations~\cite{Gore_2003, Jarzynski2006, Lechner2007, Schmiedl2007, Geiger2010, Seifert2012}.

\begin{figure*}[ht]
        \centering
        \includegraphics[width=\linewidth]{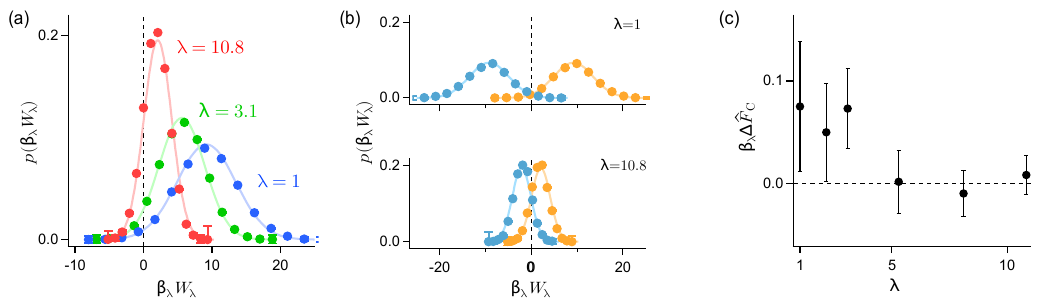}
        \caption{Experimentally rescaling the Langevin clock alters work statistics and improves free-energy estimates.
        (a) The scaled reduced work distributions $p(W)$ for different scaling factor $\lambda$. At larger $\lambda$, distributions are narrower, and closer to $\Delta F=0$ (denoted by the dashed line).  The solid lines are analytical predictions. (b) Forward (blue) and time-reversed (yellow) work distributions $p_\textrm{F}(W_\lambda),p_\textrm{R}(-W_\lambda)$. The Crooks free-energy estimator is obtained from their intersection. At higher $\lambda$, the distributions have more overlap and intersect closer to their mean values. (c) Increasing the scaling factor $\lambda$ improves the estimation of free-energy change $\Delta\hat{F}_\textrm{C}$, as obtained using the Crooks relation (Eq.~\ref{eq:Crooks_rel}). The error-bars represent standard deviation obtained from Monte-Carlo simulation. The dashed line indicates equilibrium free-energy change $\Delta F=0$.}
         \label{fig_w_delf}
\end{figure*}

    To confirm this prediction experimentally, we note that the work done making an arbitrary transformation of this system is~\cite{ peliti2021,Seifert2025}
    \begin{equation}
        W_\lambda = \int_{0}^{t_\textrm{f}}\dd t \pdv{U_\lambda(x,c(t))}{t}.
    \end{equation}
    The total work for each trial is computed as
    \begin{equation}
         W_{\lambda,i} = \sum_{n=0}^{t_\textrm{f}/\Delta t}U_\lambda(x_n,c(t_n))-U_\lambda(x_n,c(t_{n-1})).
    \end{equation}
    From $N$ repeated trials, we get a distribution of work values $W_{\lambda,i}$.
    
    The effect of the clock-rescaling parameter $\lambda$ on  work statistics is shown in Fig.~\ref{fig_w_delf}(a). With increasing $\lambda$, the work fluctuation reduces and the distribution of work becomes narrower.  The solid line represents the predicted work distributions, which are Gaussian, with mean
    \begin{equation}
        \beta_\lambda \langle W_\lambda\rangle = \frac{\tilde{c_\textrm{f}}^2}{\lambda^2 \tilde{t_\textrm{f}}^2} \left( \lambda \tilde{t_\textrm{f}} + \exp({-\lambda \tilde{t_\textrm{f}}}) - 1 \right)
        \label{eqn:w_tf}
    \end{equation}
    and variance $\sigma_W^2 = 2k_\textrm{B}T\langle W_\lambda\rangle$. Again, the rescaling of time $t\rightarrow\lambda t$ 
    reduces the effective work.
    At larger $\lambda$, the mean work approaches the equilibrium free-energy change $\Delta F=0$. This trend is demonstrated in Fig.~\ref{fig_w_delf}(a), where the mean work $\langle W \rangle$ decreases toward $\Delta F$ with increasing $\lambda$, in good agreement with Eq.~\eqref{eqn:w_tf}, whose predictions for the Gaussian distributions are shown as smooth  curves.  

    The Crooks identity relates the free-energy change to the distribution of forward (F), and reverse (R) protocol work values in a nonequilibrium transformation~\cite{Crooks1999},
    \begin{equation}
        \frac{p_\textrm{F}(W)}{p_\textrm{R}(-W)} = \exp \left(\beta W-\Delta F\right),
        \label{eq:Crooks_rel}
    \end{equation}
    and the equilibrium free energy is obtained from the intersection $W^\textrm{C}$,
    \begin{equation}
        \begin{aligned}
            p_\textrm{F}(W^\textrm{C}) &= p_\textrm{R}(- W^\textrm{C})\\
            \implies \Delta \hat{F}_\textrm{C} &= W^\textrm{C}.
        \end{aligned}
    \end{equation}
    
    We divide our trials into two equally numbered groups and treat them as forward (F) and time-reversed (R) realizations.  The probability density of work values, $p_\textrm{F}(W_\lambda),p_\textrm{R}(-W_\lambda)$ follow Gaussian distributions, as seen in Fig.~\ref{fig_w_delf}(b).  From the mean and variance of these distributions, we compute the intersection $W^\textrm{C}_{\lambda}\equiv \Delta \hat{F}_\textrm{C}$ of two corresponding Gaussian functions (see Supplementary Material Sec.~V). Because of this Gaussian assumption, the Crooks relation leads to more precise and accurate estimates of the equilibrium free-energy change with increasing Langevin-clock rescaling (see Fig.~\ref{fig_w_delf}(c)).  For a similar analysis of free-energy difference based on the Jarzynski relation, see Sec.~V in Supplementary Material~\cite{SI}.
    
    To compute the statistical uncertainty of our free-energy estimator, we perform Monte-Carlo simulation of our protocol for 100 repetitions of our experimental time.  We see that the uncertainty reduces with increasing $\lambda$ (See Sec.~V, VI in Supplementary Material~\cite{SI}). Thus, the dynamics scaling method reduces dissipated work and in effect suppresses estimator variance, enabling more reliable free-energy estimates from the same length of trajectories.

\textit{Discussion---}Our approach has advantages over other possible control strategies, including shortcuts to adiabaticity and optimal control~\cite{Sekimoto2010,Odelin2019,Odelin2023,Blaber2023,alvarado2025}.  Both use bespoke protocols designed for a particular system with particular parameters to speed up state transformations~\cite{Martinez2016,Menu2022}.  They are inherently task specific, requiring a different protocol for each potential landscape. Our method instead trades optimality for universality. Our Langevin-clock rescaling results in better statistics than the optimal protocol, which minimizes both work and uncertainty of free-energy estimate~\cite{Schmiedl2007,Geiger2010,Whitelam2025}. By applying simple, global modifications---uniformly scaling deterministic forces and adding proportional broadband noise---we accelerate dynamics across a large class of potentials without the need to solve control problems for every new trajectory and without the need for an accurate model of the system under study. This establishes a strategy where noise acts as a functional resource for general mesoscopic control.

These capabilities have immediate implications for thermodynamic computing, where computational throughput is often limited by the time required for stochastic devices to equilibrate within an energy landscape~\cite{Alemi2018,Hylton2020,Wimsatt2021,Boyd2022,Aadit2022,Coles2023,Misra2023,Wolpert2024, aifer2024,Melanson2025,Duffield2025, Whitelam2026}. By rescaling the Langevin clock, we effectively increase the mobility of the computational degrees of freedom~\cite{Whitelam2025a}. As demonstrated by the rapid convergence of position distributions in Fig.~\ref{fig_px_xt}(a) and the reduced variance of Jarzynski and Crooks free-energy estimators in Fig.~\ref{fig_w_delf}(c), increasing the scaling factor $\lambda$ accelerates relaxation and improves thermodynamic inference without altering the equilibrium distribution of the computer. Our method leaves the signal-to-noise ratio unchanged (here, the SNR is the ratio of thermal  motion in the optical trap to measurement noise of particle position).  By contrast, naively scaling only the forces (Fig.~\ref{fig_scheme}(b)) will reduce the SNR.  In applications such as thermodynamic computing, lower SNR leads to reduced precision or increased effective sampling time to reach a target confidence~\cite{Melanson2025}.

Similar scaling of both deterministic and fluctuating forces can be implemented in diverse experimental setups. For instance, in mesoscopic electronic systems, added current noise would achieve an equivalent speedup~\cite{Melanson2025, GUARCELLO2021}. Quantum reservoir engineering suggests that Hamiltonian rescaling combined with controlled dissipation might similarly accelerate convergence to quantum steady states, an intriguing direction for future inquiry~\cite{ Ivander2023, Sgroi2025}.

    Our approach addresses the fundamental challenge of accelerating relaxation in mesoscopic systems by redefining noise as a functional resource rather than a nuisance. The simultaneous scaling of deterministic and fluctuating forces enables a speed-up of Langevin dynamics, yielding both higher computational throughput and improved precision in nonequilibrium free-energy estimation. By harnessing the interplay between fluctuations and dissipation, this strategy offers a universal alternative to task-specific control protocols, applicable to diverse fields ranging from thermodynamic computing to single-molecule manipulation.

{\em Acknowledgments}---We thank David Sivak for catalyzing this collaboration.  S.W. performed work at the Molecular Foundry, supported by the Office of Science, Office of Basic Energy Sciences, of the U.S. Department of Energy under Contract No. DE-AC02-05CH11231, and was partly supported by US DOE Office of Science Scientific User Facilities AI/ML project ``A digital twin for spatiotemporally resolved experiments''. JB was supported by a Discovery Grant from the National Science and Engineering Research Council (NSERC), Canada.

\nocite{Kumar2018,Bryan2022,Martinez2013, Saha2023, Message2025, Chupeau2018, Loi2008, Petrelli2020, DiBello2024,BergSoerensen2004, Jarzynski1997}

%

\clearpage
\onecolumngrid

\setcounter{equation}{0}
\setcounter{figure}{0}
\setcounter{table}{0}
\renewcommand{\theequation}{S\arabic{equation}}
\renewcommand{\thefigure}{S\arabic{figure}}
\renewcommand{\thetable}{S\arabic{table}}
\renewcommand{\theHequation}{S\arabic{equation}}
\renewcommand{\theHfigure}{S\arabic{figure}}
\renewcommand{\theHtable}{S\arabic{table}}

\begin{center}
\Large \textbf{Supplementary Information for\\
Adding noise and scaling forces to speed up the Langevin clock}
\end{center}

\author{Prithviraj Basak}
\affiliation{Department of Physics, Simon Fraser University, Burnaby, British Columbia V5A 1S6, Canada}

\author{Stephen Whitelam}
\affiliation{Molecular Foundry, Lawrence Berkeley National Laboratory, 1 Cyclotron Road, Berkeley, CA 94720, USA}

\author{John Bechhoefer}
\affiliation{Department of Physics, Simon Fraser University, Burnaby, British Columbia V5A 1S6, Canada}

\maketitle

\section{Experimental Details}
    \emph{Apparatus---}The feedback optical tweezer setup is described in detail in Refs.~\cite{Kumar2018,Bryan2022}. Here, we briefly review the key components. As shown below in Fig.~\ref{SI_fig_apparatus}, a 532-nm, continuous-wave, diode-pumped laser (HUBNER Photonics, Cobolt Samba, 1.5~W) serves as the trapping light source, coupled with a custom microscope mounted on a vibration-isolation table (Melles Griot). The trapping beam passes through acousto-optic deflectors (DTSXY-250-532, AA Opto Electronic) to enable XY-plane deflection, while we control the beam angle and intensity with analog voltage oscillators (DFRA10Y-B-0-60.90, AA Opto Electronic). A telescopic lens system increases the diameter of the beam to overfill the back aperture of the trapping objective (MO1, Olympus 60X, UPlanSApo, NA = 1.2), allowing us to trap 1.5~µm silica beads in an aqueous solution.
    \begin{figure}[htb]
        \centering
        \includegraphics[ ]{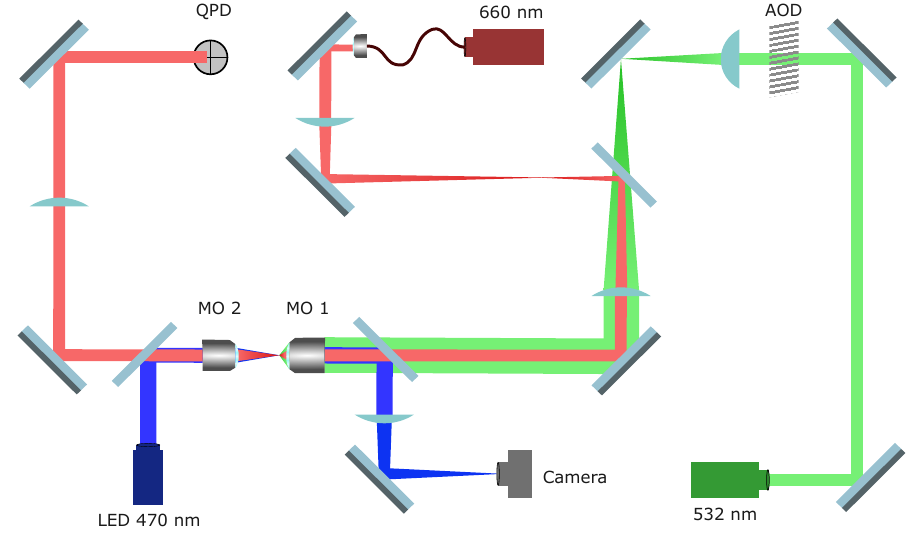}
        \caption{Schematic diagram of the experimental apparatus. The 532~nm green laser is focused by microscope objective MO1 and traps a silica sphere.  The forward scattered light from the 660~nm red laser is imaged on a quadrant photodiode (QPD) for position detection. Blue light from a 470nm LED is used to image the trapped sphere on the camera. An FPGA records the position data from the QPD and controls the deflection of the green laser using the AOD.}
        \label{SI_fig_apparatus}
    \end{figure}

    We focus the detection beam (662~nm, Thorlabs, LP660-SF50) through the same trapping microscope objective and adjust it with a 4f relay lens system. A low-numerical-aperture objective (MO2) (40X, NA = 0.4) collects the forward-scattered light from the trapped bead and separates it from the trapping beam using a high-pass filter (cut-off 585~nm, Edmund Optics).The detection beam then reaches a quadrant photodiode (QPD, First Sensor, QP50-6-18u-SD2).
    
    For imaging, we illuminate the trapped beads with a blue LED (470~nm, Thorlabs, M470D3). The light reflects off a dichroic long-pass filter (cut-off 585~nm, Edmund Optics) and enters through MO2 to the sample solution.  After MO1 collects the light, another dichroic long-pass filter (cut-off 500~nm, Edmund Optics) separates it from trapping and detection beam before reaching a camera (FLIR BFS-U3-04S2M-CS). We acquire data using a LabView program, which employs a field-programmable gate array (FPGA, National Instruments, NI PCIe-7857) to convert analog signals from the QPDs into digital position signals. This system processes the signals to implement the desired  control loop and sends the control signals to the AODs.

\begin{figure}[htb]
        \centering
        \includegraphics[ ]{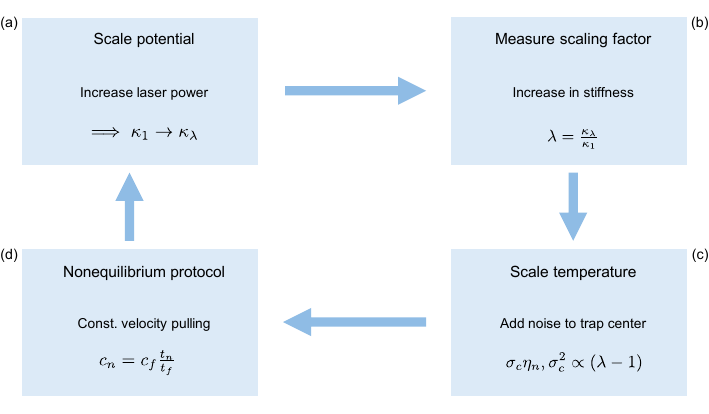}
        \caption{Experimental flowchart for the spatial-temporal scaling protocol. (a) We first determine the baseline trap stiffness ($\kappa_1$) and diffusion constant at $\lambda = 1$. We then increase the trap laser power via an acousto-optic deflector (AOD) and (b) quantify the realized scaling factor $\lambda = \kappa_\lambda/\kappa_1$ by measuring the scaled stiffness $\kappa_\lambda$. (c) Using this $\lambda$, we calculate the requisite variance for the trap-center fluctuations using Eq.~7 to appropriately scale the effective temperature. (d) After scaled stationary conditions are reached, the nonequilibrium transformation is executed by translating the trap at a constant velocity.}
        \label{SI_fig_flowchart}
    \end{figure}

    \emph{Scaling---}We realized the speed-up method at different scaling factor values $\lambda$, ranging from 1 to 10.8.  We implement the scaling in several steps, as illustrated in Fig.~\ref{SI_fig_flowchart}.  First, we increase the trapping laser power and estimate the achieved scaling factor from the measured stiffness increase as $\lambda = \kappa_\lambda / \kappa_1$.  Next, we add noise to our system using optical forces to raise the effective temperature to $T_\textrm{eff} = \lambda T$. Finally, the trap is translated at a constant velocity to perform the nonequilibrium protocol. Before each trial, the system equilibrates for a time $5\tau_{\textrm{r}}$.
    
    \emph{Parameters---}We sample the bead's position and apply the control at a time-step of $\Delta t = 20$~\textmu s. Without any scaling, at the lowest laser power where the trap is stable, we find a diffusion constant $D=0.31$~\textmu m$^2$/s and trap stiffness $\kappa_1=4.2$~pN/\textmu m. Thus, for the nonequilibrium pulling protocol, we use the trap displacement $c_\textrm{f} = 5\times \sigma_x = 155.1~ \textrm{nm}$, and protocol time $t_\textrm{f} = \tau_{\textrm{r}} = 0.0031~\textrm{s}$. We repeated the constant velocity protocol $\approx$ 23,000 times for each scaling factor.

\newpage

\section{Insights on faster relaxation}
\begin{figure}[htb]
        \centering
        \includegraphics[]{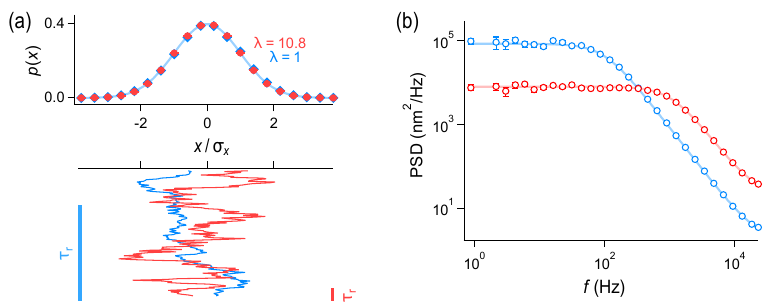}
        \caption{Faster relaxation with invariant stationary distributions under Langevin dynamics scaling in experiment. The probability density of measured position $p(x)$ for different scaling factors ($\lambda = 1$ and $10.8$) in a static harmonic potential follows the same Boltzmann distribution $p(x) = \frac{1}{\mathcal{Z}}\exp(-\frac{\kappa_1x^2}{2k_\textrm{B}T})$, where $\mathcal{Z}$ is the normalization factor. Despite this identical spatial distribution, the dynamics relax faster for higher scaling factors. This relaxation speed-up is evident from the increased corner frequency $f_\textrm{c} = \frac{1}{2 \pi\tau_\textrm{r}}$ seen in their respective power spectral densities (PSD).}
        \label{SI_fig_speedup}
    \end{figure}

    The physical mechanism behind the relaxation speed-up becomes evident when comparing the power spectral densities (PSD) of the bead's position. In Fig.~\ref{SI_fig_speedup}, we examine the system with a stationary trap, comparing the unscaled dynamics ($\lambda = 1$) against the maximum scaling achieved in our experiment ($\lambda = 10.8$). While the stationary probability distribution $p(x)$ remains identical in both cases, the trajectory $x(t)$ for the scaled system relaxes significantly faster. As the red markers and fit curve in Fig.~\ref{SI_fig_speedup}(b) show, this enhanced relaxation rate increases the corner frequency $f_\textrm{c} = 1/(2\pi \tau_\textrm{r})$ of the PSD.  The power-spectra fits in Fig.~\ref{SI_fig_speedup} are to the aliased Lorentzian~\cite{BergSoerensen2004}
    \begin{equation}
    \langle |\hat{x}(f)|^2 / t_\textrm{tot} \rangle = \frac{(\Delta_2)^2 \Delta t}{1 + a^2 - 2a \cos(2\pi f \Delta t)},
\end{equation}
    where $a= e^{-\Delta t/\tau_\textrm{r}}, \Delta_2 = \sqrt{(1-a^2)/\kappa_\lambda}$. The parameter $t_\textrm{tot}$ is the total duration of the trajectory from which PSD is calculated.
    
    To understand this theoretically, it is instructive to rescale time as $t \to t/\lambda$ in Eq.~3 to obtain
    \beq
    \label{eff2}
    \dot{x} = -\mu \frac{\partial U(x,t/\lambda)}{\partial x} + \sqrt{2 k_\textrm{B}T \mu} \, \eta(t).
    \eeq
    This modified system evolves over the extended time interval $t \in [0,\lambda t_\textrm{f}]$. For an undriven system where $U(x,t) = U(x)$, the probability distribution $P_\lambda(x,t/\lambda)$ of the modified system at time $t/\lambda$ is identical to the distribution $p(x,t)$ of the original system at time $t$. If the system is driven out of equilibrium by a time-dependent protocol $U(x,t)$, the modified dynamics are equivalent to the original system subjected to a protocol applied $\lambda$ times more slowly and for $\lambda$ times longer. In this sense, increasing the scaling factor effectively speeds up the Langevin clock by a factor of $\lambda$. 
            
    Crucially, this clock rescaling leaves the equilibrium probability distribution of $x$ completely unchanged. To see this, we can rewrite Eq.~3 to explicitly show the scaled physical parameters:
    \beq
    \label{eff3}
    \dot{x} = -\mu \frac{\partial U_\lambda(x,t)}{\partial x} + \sqrt{2 k_\textrm{B} T_\lambda \mu} \, \eta(t).
    \eeq
    This form highlights the rescaled potential $U_\lambda \equiv \lambda U$ and the effective temperature $T_\lambda \equiv \lambda T$. Consequently, the rescaled reciprocal temperature is $\beta_\lambda \equiv \beta/\lambda$, where $\beta \equiv 1/(k_\textrm{B}T)$. Because both the potential and the temperature scale linearly with $\lambda$, the Boltzmann distribution of any configuration $x$ remains invariant, i.e., $\e^{-\beta_\lambda U_\lambda} = \e^{-\beta U}$. Thus, the equilibrium distribution is preserved such that $P_\lambda(x) = p(x)$, consistent with the experimental observations in Fig.~\ref{SI_fig_speedup}(a).

\section{Improvements in the work statistics}
The work distributions demonstrated a reduction in mean and consequently variance for larger scaling factors $\lambda$ (Fig.~3(a)). For the linear translation protocol, the dimensionless mean work $\beta_\lambda \langle W_\lambda(t)\rangle$ evolves analytically according to
\begin{equation}
    \beta_\lambda \langle W_\lambda(t)\rangle = \frac{\tilde{c}_f^2}{\lambda^2 \tilde{t}_f^2} \left( \lambda \tilde{t} + \exp({-\lambda \tilde{t}}) - 1 \right),
    \label{eqn:w_t}
\end{equation}
where we recall the dimensionless parameters defined in the main text: the total translation distance $\tilde{c}_\textrm{f}= c_\textrm{f}/\sigma_x$, the protocol duration $\tilde{t}_\textrm{f}= t_\textrm{f}/\tau_\textrm{r}$, and the time $\tilde{t} = t/\tau_\textrm{r}$.
    \begin{figure}[htb]
        \centering
        \includegraphics[]{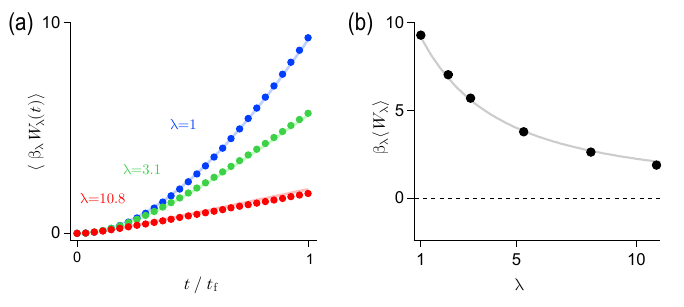}
        \caption{Reduction of scaled mean work and improved convergence of the Jarzynski estimator. (a) Time evolution of the scaled mean work $\beta_\lambda \langle W(t)\rangle$ over the protocol duration as measured in experiment. Solid lines indicate analytical predictions derived from Eq.~\eqref{eqn:w_t}. (b) Total scaled mean work $\beta_\lambda \langle W \rangle$ as a function of the scaling factor $\lambda$, showing excellent agreement with the theoretical prediction (solid line) given by Eq. 11.}
        \label{SI_fig_work}
    \end{figure}

With increased Langevin-clock speed up, the bead relaxes closer to the equilibrium position compared to the unscaled dynamics (Fig.~2).  As a result, the scaled work grows more slowly with time at higher $\lambda$, as seen in Fig.~\ref{SI_fig_work}(a). Evaluated at the end of the protocol, the total scaled mean work decreases monotonically as $\lambda$ increases (Fig.~\ref{SI_fig_work}(b)).
    
\section{Crooks estimator from the intersection of two Gaussian distributions}    
We evaluate the free-energy change via the Crooks fluctuation theorem by identifying the crossing point of the forward (F) and reverse (R) work distributions. Leveraging the empirically verified Gaussian nature of these distributions, we analytically determine their intersection $W^\textrm{C}$ using the sample means and variances, $\{\langle W_\textrm{F} \rangle,\sigma^2_\textrm{F}\}$ and $\{\langle W_\textrm{R} \rangle,\sigma^2_\textrm{R}\}$. This crossing point is a root of the quadratic equation $AX^2 + B X + C = 0$, with coefficients defined as
\begin{align}
    A &= \frac{1}{\sigma_\textrm{F}^2} - \frac{1}{\sigma_\textrm{R}^2}, \nonumber \\
    B &= 2 \left( \frac{\langle W_\textrm{R} \rangle}{\sigma_\textrm{R}^2} - \frac{\langle W_\textrm{F} \rangle}{\sigma_\textrm{F}^2} \right), \nonumber \\
    C &= \frac{\langle W_\textrm{F} \rangle^2}{\sigma_\textrm{F}^2} - \frac{\langle W_\textrm{R} \rangle^2}{\sigma_\textrm{R}^2} - 2\ln\left(\frac{\sigma_\textrm{F}}{\sigma_\textrm{R}}\right).
\end{align}
The appropriate root $W^\textrm{C}$, which represents the Crooks estimator, is uniquely chosen as the value located in the interval between the two means $\langle W_\textrm{F} \rangle$ and $\langle W_\textrm{R} \rangle$.

\begin{figure}[htb]
    \centering
    \includegraphics[]{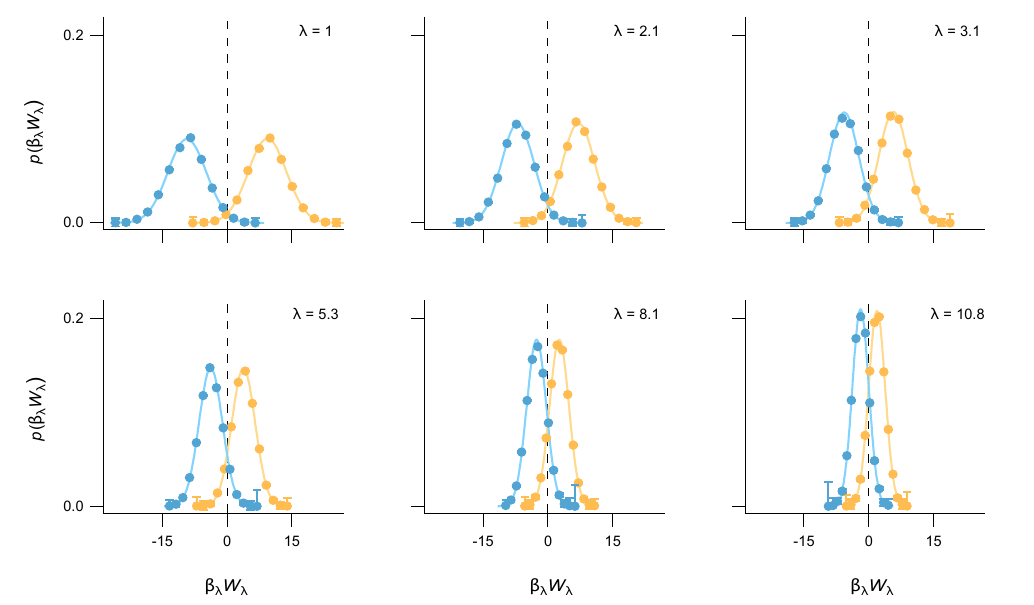}
    \caption{Crooks estimator from Gaussian intersections of measured work distributions. Probability density of the forward and reverse protocol work values $\beta_\lambda W_\lambda$. The solid curves represent ideal Gaussian functions parameterized by the sample means and variances of the empirical work values. The Crooks free-energy estimate is extracted analytically from the intersection of these two constructed distributions. The dashed line indicates the equilibrium free-energy difference.}
    \label{SI_fig_crooks}
\end{figure}

As illustrated in Fig.~\ref{SI_fig_crooks}, the empirical work values are consistent with their analytical Gaussian predictions. Extracting the free energy from the intersection of the analytical curves that follow their measured mean and variance (not the predicted values), rather than directly from the overlapping histogram bins, significantly mitigates finite-sampling errors in the extreme statistical tails where the forward and reverse distributions cross.  In other words, knowing \textit{a priori} that the probability densities are Gaussian increases greatly the accuracy of the free-energy estimation.

\section{Jarzynski estimator for free-energy difference}

      In transformations where only the forward trajectory is available, the Jarzynski relation~\cite{Jarzynski1997} is a useful alternative to the Crooks relation. The free-energy estimator based on the Jarzynski relation is computed as
    \begin{equation}
    \begin{aligned}
            \beta_\lambda\Delta \hat{F_\textrm{J}} 
                &= - \ln \left\langle \exp({-\beta_\lambda W}) \right\rangle \\
                &= \ln N - \ln \left( \sum_{i=1}^{N} \exp({-\beta_\lambda  W_{\lambda,i}}) \right).
    \end{aligned}
    \label{eq:Jarz}
    \end{equation} 
    \begin{figure}[htb]
        \centering
        \includegraphics[]{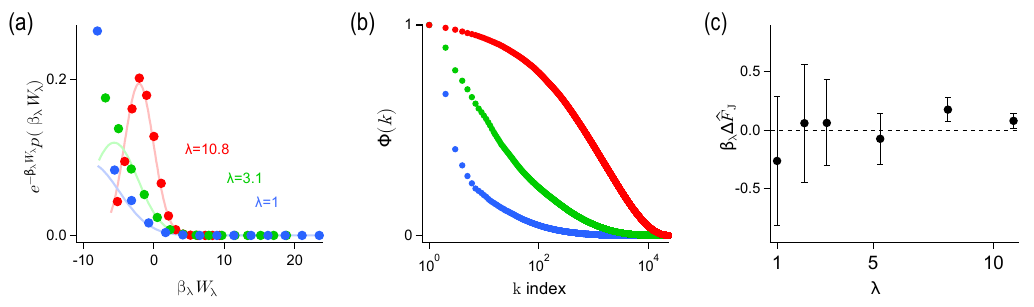}
        \caption{Convergence of the Jarzynski estimator with increasing scaling factor. (a, b) The fractional contribution of lowest-work trajectories to the exponentiated sum of the Jarzynski estimator. The cumulative fraction is defined as $\Phi(k) = \sum_{i=k}^N \exp(-\beta_\lambda W_{\lambda i}) / \sum_{i=1}^N \exp(-\beta_\lambda W_{\lambda i})$, with work values $W_{\lambda i}$ sorted in descending order. The disproportionate dominance of rare events is heavily suppressed at larger $\lambda$. (c) Free-energy difference calculated using Jarzynski estimator in Eq.~\ref{eq:Jarz} from the experimental data. The error bars are standard deviations of the Jarzynski estimator calculated from Monte-Carlo simulation of 100 times the experimental duration. The dashed line indicates the equilibrium free-energy change.}
        \label{SI_fig_Jarz}
    \end{figure}

    For the unscaled system, the exponential average $\langle \exp(-\beta_\lambda W_\lambda) \rangle$ in the Jarzynski estimator is dominated by rare trajectories in the left tail of the work distribution (Figs.~\ref{SI_fig_Jarz}(a) and \ref{SI_fig_Jarz}(b). However, in our method, with increasing scaling factor, these rare values occur with smaller probability and we sample the work statistics more effectively. By increasing $\lambda$, the distribution shifts closer to the reversible work, making the exponential average less dependent on extreme events. To quantify this effect, we evaluate the fractional contribution of the lowest-work trajectories to the total Jarzynski sum, defined as $\Phi(k) = \sum_{i=k}^N \exp(-\beta_\lambda W_{\lambda i}) / \sum_{i=1}^N \exp(-\beta_\lambda W_{\lambda i})$, where the observed work values $W_{\lambda i}$ are sorted in descending order. As demonstrated in Figs.~\ref{SI_fig_Jarz}(a) and \ref{SI_fig_Jarz}(b), the sum for the unscaled dynamics ($\lambda=1$) is largely determined by the lowest-work values. For larger $\lambda$, the contribution from these rare trajectories is suppressed, allowing the Jarzynski estimator $\Delta \hat{F_\textrm{J}}$ to converge better with the same number of trials (Fig~\ref{SI_fig_Jarz}(c)). 
    
\begin{figure}[htb]
        \centering
        \includegraphics[]{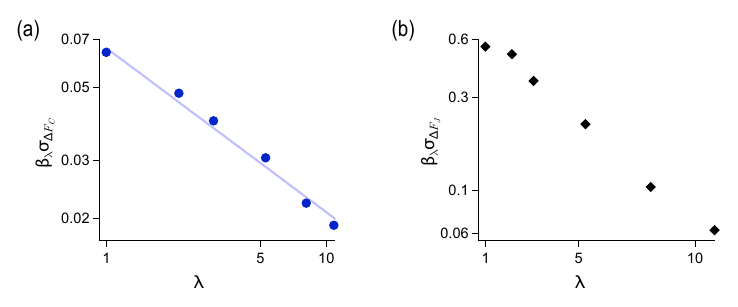}
        \caption{Reduction in the free-energy estimator uncertainty with increasing speed up. (a) Log-log plot of the standard deviation of Crooks estimator, computed by Monte-Carlo simulation.  Solid line is $\sim \lambda^{-1/2}$.  (b) Log-linear plot shows an exponential reduction of the uncertainty of Jarzynski estimator, also calculated by Monte-Carlo simulation. }
        \label{SI_fig_delF_sd}
    \end{figure}

    \emph{Uncertainty in the free-energy difference---}We quantify the uncertainty in the Jarzynski and Crooks estimators using Monte Carlo simulations of the nonequilibrium protocol, generating a statistical ensemble 100 times the experimental sample size. The uncertainties, $\sigma_{\Delta F_\textrm{J}}$ and $\sigma_{\Delta F_\textrm{C}}$, correspond to the standard deviations of the respective estimators across these repetitions (Fig.~\ref{SI_fig_delF_sd}). The uncertainty in the Jarzynski estimator decreases exponentially with the scaling factor $\lambda$(Fig.~\ref{SI_fig_delF_sd}(b)). The uncertainty in the Crooks estimator, while initially lower, decreases more gradually (Fig.~\ref{SI_fig_delF_sd}(a)).  

    To understand the observed power-law decrease in standard deviation for the Crooks estimator, we consider the following simple picture:  Two Gaussian work distributions with identical standard deviation $\sigma_W$ and means $\langle W_\textrm{F} \rangle, \langle W_\textrm{R} \rangle $, intersect. The Crooks estimator is then $\Delta F_\textrm{C} = (\langle W_\textrm{F} \rangle +\langle W_\textrm{R} \rangle )/2$. As shown in Eq.~11 and Fig.~\ref{SI_fig_work}(b), both the scaled mean work $\beta_\lambda\langle W_\lambda \rangle $ and consequently the variance scale inversely with $\lambda$. Hence, the uncertainty in the Crooks estimator divided by the standard deviation is
    \begin{equation}
    \begin{aligned}
        \frac{\beta_\lambda \delta \Delta F_\textrm{C}}{\beta_\lambda \sigma_W} &\propto \frac{\beta_\lambda (\langle W_\textrm{F} \rangle +\langle W_\textrm{R} \rangle )}
        {\beta_\lambda \sigma_W} \propto \frac{1/\lambda}{1/\sqrt{\lambda}},\\
        \frac{\delta \Delta F_\textrm{C}}{\sigma_W} &\propto \frac{1}{\sqrt{\lambda}}.
    \end{aligned}
    \end{equation}
    Similarly, the Jarzynski estimator calculated from the exponentiated work, decreases exponentially with increasing Langevin-clock speed-up.

\section{Experimental Limitations}
    \emph{Drift---}To determine whether spatial drift in our experimental setup impacts our thermodynamic measurements, we calculate its theoretical contribution to the dissipated work. If we assume a constant drift velocity $v_\textrm{d}$, the work dissipated over a protocol duration $\tau_\textrm{f}$ is $\Delta W_\textrm{d} = \gamma v_\textrm{d}^2 \tau_\textrm{f}$, where $\gamma$ is the viscous friction coefficient of the bead. We use the Stokes-Einstein relation $\beta \gamma = 1/D = \tau_\textrm{r}/\sigma_x^2$ (where $D$ is the diffusion constant) and noting that our protocol duration equals the trap relaxation time ($\tau_\textrm{f} = \tau_\textrm{r}$), this drift-induced work simplifies to:
    \begin{equation}
        \beta \Delta W_\textrm{d} = \frac{v_\textrm{d}^2 \tau_\textrm{r}}{D} =  v_\textrm{d}^2 \frac{ \tau_\textrm{r}^2}{\sigma_x^2}.
    \end{equation}
    We compare the dissipation due to drift to the mean work $\langle W \rangle$ generated by the linear translation protocol. For a trap moving at a constant protocol velocity $v_c$ over the same duration, the expected dimensionless mean work is:
    \begin{equation}
        \beta \langle W \rangle \approx v_c^2 \frac{ \tau_\textrm{r}^2}{\sigma_x^2}.
    \end{equation}
    
    In our experimental setup, the typical average drift velocity is $v_\textrm{d} \leq 10$~nm/s. By contrast, the trap is pulled at a velocity $v_c = 5\sigma_x/\tau_\textrm{r} = 51$~\textmu m/s. This yields a velocity ratio of $v_d/v_c \approx 10^{-4}$. Consequently, the relative error in the work due to drift scales as the square of this ratio:
    \begin{equation}
        \frac{\Delta W_\textrm{d}}{\langle W \rangle}  = \frac{v_\textrm{d}^2}{v_c^2} \approx 10^{-8}.
    \end{equation}
    Because this contribution is eight orders of magnitude smaller than the protocol work, we conclude that the inherent experimental drift has a negligible effect on our measured work statistics and subsequent free-energy calculations.

\begin{figure}[htb]
        \centering
        \includegraphics[ ]{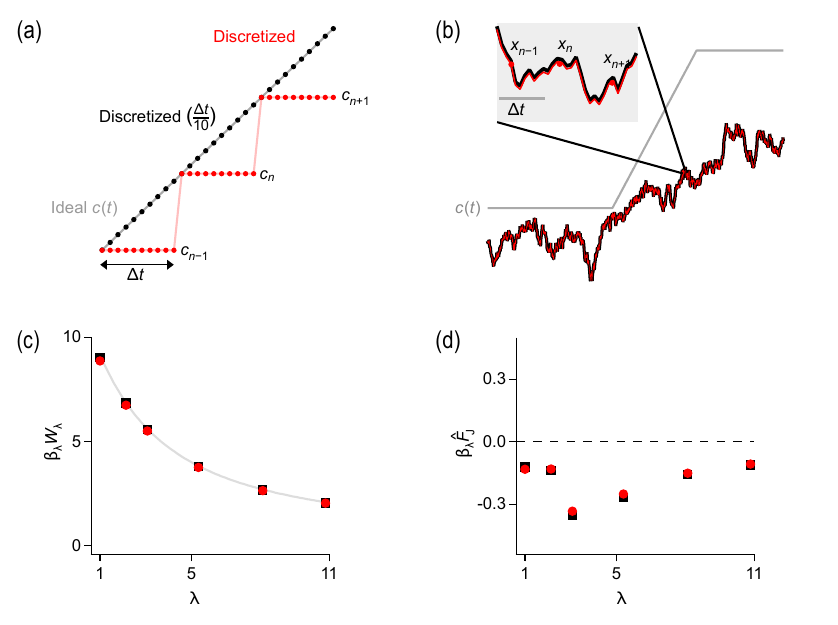}
        \caption{Numerical simulation of the effect of discrete control on the bead's trajectory, work, and free-energy estimation. (a) Illustration of different types of trap-motion simulated using time step $\tfrac{\Delta t}{10}$---updated at every time-step (black), updated at 10$\times$ time-step $=\Delta t$ (Red). (b) Comparison of trajectories generated by the two types of discrete trap update. (c) From the simulated trajectories, scaled mean work is obtained. The solid line represents the analytical prediction based on Eq.~11. (d) Scaled Jarzynski free-energy estimator evaluated from these same simulations. The dashed line indicates the equilibrium free-energy difference. }
        \label{SI_fig_limit}
    \end{figure}

     \emph{Discretized control---}In our experiment, the trap is updated once every sampling time $\Delta t = 2\cdot10^{-5}$~s.  Between two successive updates, the trap position is approximately fixed, leading to a staircase-like motion (see Fig.~\ref{SI_fig_limit}).  We investigated and found there is no apparent difference in the dynamics of the bead compared to a truly continuous constant-velocity motion.  For $\lambda = 1,$ we have $ \tau_\textrm{r}/\Delta t\approx150$, which updates the trap much faster than the relaxation dynamics of the bead. Even in the case of $\lambda=10.8$, we have $ \tau_\textrm{r}/\lambda \Delta t\geq 10$.

      \emph{Numerical simulation---}To evaluate the robustness of our system against the effect of discrete trap motion, we perform a Monte-Carlo simulation of our linear protocol using the experimental parameters. As illustrated in Fig.~\ref{SI_fig_limit}(a), we simulate the trap-motion and bead-dynamics with a faster time-step $t_\textrm{s} = \tfrac{\Delta t}{10}=2$~\textmu s, compared to the experimental sampling time.  We use the following propagator for the discretized Langevin dynamics,
    \begin{equation}
        x_n = (a (1 - c_n) + c_n ) x_{n-1} + \Delta_2 \eta_n + \sigma_c\eta^c_{n},
        \label{eq:discrete-dyn}
    \end{equation}
    where $a= e^{-\Delta t/\tau_\textrm{r}}, \Delta_2 = \sqrt{(1-a^2)/\kappa_\lambda}$, and $\sigma_c$ is given in Eq.~7. Here, $\eta_n$ and $\eta^c_{n}$ are independent pseudorandom numbers for the thermal and added noise respectively, with $\av{\eta_n}=\av{\eta^c_n}=0$ and $\av{\eta_n \eta_m}=\av{\eta^c_n \eta^c_m}=\delta_{nm}$ and $\av{\eta^c_n \eta_m}=0$.  We recall that $c_n$ is the trap position at time-step $n$.

    We implement two distinct update-strategies for the trap-motion $c_n$---one updated at every finite time-step $t_\textrm{s}$, and the other updated at  the experimental time resolution $\Delta t = 10 t_\textrm{s}$ (see Fig.~\ref{SI_fig_limit}(a)). As demonstrated in Fig.~\ref{SI_fig_limit}(b), the underlying bead dynamics are largely insensitive to these variations in trap motion. Consequently, the discrete updating introduces no systematic bias into the measured thermodynamic quantities: the mean work and Jarzynski free-energy estimates remain consistent across all three realizations (Figs.~\ref{SI_fig_limit}(c) and (d)).
    
    To obtain the uncertainty in our experimental measurements, we combine 100 repetitions of each scaling factor realization, where each repetition contains $2\times 10^4$ trials.  
    From these 100 repetitions, we find the standard deviation of the Jarzynski and the Crooks estimator as our statistical uncertainty.

    \emph{Clipping of maximum force---}Optical tweezers are limited by the finite escape force of the trapped bead. When introducing additional fluctuations via trap motion, we must ensure that the maximum applied force, $F^{\textrm{max}}_\lambda$, does not exceed the escape force, $F^{\textrm{esc}}_\lambda$. Because we increase the laser power $\mathcal{P}_\lambda$ to scale the deterministic force, the escape force scales linearly with the scaling factor $\lambda$:
    \begin{equation}
        F^{\textrm{esc}}_\lambda \propto \mathcal{P}_\lambda \propto \lambda.
    \end{equation}
    Concurrently, the maximum force exerted on the bead scales at the exact same rate:
    \begin{equation}
        F^{\textrm{max}}_\lambda \propto \kappa_\lambda\sigma_x \propto \lambda.
    \end{equation}
    Consequently, the ratio of the maximum applied force to the escape force remains constant, ensuring that the force trajectories are not artificially clipped at higher scaling factors.
    
    \emph{Effective temperature---}One way to increase the effective temperature $T_\lambda$ of the trapped bead is to add a non-equilibrium force~\cite{Martinez2013,  Saha2023,Message2025}.  This is often achieved by using electrical forces, as the beads carry static charge.  Here, instead of external forces, we apply optical forces by adding random fluctuations to the position of the trap, resulting in an added white noise force on top of the thermal noise from the bath~\cite{Chupeau2018}. Operationally, this method of adding noise force is equivalent to increasing the effective temperature, provided that the bandwidth of the injected noise exceeds the bandwidth of the dynamics given by the system's corner frequency~\cite{Loi2008, Petrelli2020, Saha2023, DiBello2024}. At the highest scaling factor, the corner frequency of the bead's motion $f_\textrm{c}\approx1$~kHz $\ll f_\textrm{Nyq}=25$~kHz, the bandwidth of the added noise. Here, $f_\textrm{Nyq}$ is the Nyquist frequency of the sampled signal.

\end{document}